\documentclass[journal ]{new-aiaa}
\usepackage[utf8]{inputenc}
\usepackage{textcomp}

\usepackage{graphicx}
\usepackage{amsmath}
\usepackage[version=4]{mhchem}
\usepackage{siunitx}
\usepackage{longtable,tabularx}
\usepackage{natbib}

\title{Suppression of Chaotic Motion of Tethered Satellite Systems Using Tether Length Control}

\author{F. J. T. Salazar \footnote{Postdoctoral researcher, Orbital Mechanics Department, Av. dos Astronautas, 1758; e7940@hotmail.com.} and A. F. B. A. Prado\footnote{Full researcher, Postgraduate Division, Av. dos Astronautas, 1758; antonio.prado@inpe.br; Professor, Academy of Engineering, RUDN University, Miklukho-Maklaya street 6, Moscow, Russia, 117198.}}
\affil{National Institute for Space Research (INPE), S\~ao Jos\'e dos Campos, S\~ao Paulo, Brazil, 12227-010}

\begin{document}

\maketitle

\section{Introduction}
\lettrine{T}{he} concept of a Tethered Satellite System (TSS) consists of a long cable used to couple spacecraft together as they orbit a central body (e.g. the Earth). The dynamics of a TSS was analyzed by Chobotov in 1963 \cite{Chobotov:1963} and its feasibility was tested in 1966 when the Gemini 11 space vehicle was tethered to the rocket stage Arena \cite{Odell:1966}. In later decades, several missions have already been launched to verify the TSS concept \cite{Sasaki:1987, Bortolami:1993, Gullahorn:1993, Tyc:1993} and innumerable literature were published by many different researchers to propose the TSS for space applications, such as artificial gravity, payload orbit transfer, attitude stabilization, spacecraft formation, space debris removal, and many others \cite{Chobotov:1967, Colombo:1982, Bekey:1983, Bekey:1986, Carroll:1986, Kyroudis:1988, Kumar:1992, Bekey:1997, Kumar:1997, Kurnar:1997, Kumar:2001, Kumar:2002, Williams:2002, Kumar:2004, Guerman:2008, Aslanov:2013, Aslanov:2017}. However, the limited success of the TSS-1 and TSS-1R experiments, launched in 1992 and 1996, respectively, showed that the flexibility and elasticity of the space tether lead to complex issues with the TSS \cite{Cosmo:1997}, i.e. the deployment and retrieval dynamics and control of a TSS is more complicated, and further studies and experiments should be carried out. Thus, several strategies were proposed for the control of this complex dynamical system. Rupp \cite{Rupp:1975}, Baker et al. \cite{Baker:1976}, and Bainum and Kumar \cite{Bainum:1980} were the first researchers to introduce a tension control law (TCL) to regulate the tether swing. A lot of work has been done since then, in particular on more sophisticated models that study the effect of atmospheric drag, Earth's oblateness, and electrodynamic force on the pitch and roll motion of a TSS \cite{Pelaez:2003, Pelaez:2005, Kojima:2009, Yu:2020}.     

This study focuses on attitude and control motion of two bodies (a base-satellite and a sub-satellite) connected by an inextensible and massless tether in a circular orbit under the influence of the Earth's gravitational force. The base-satellite is assumed to be far more heavier than the sub-satellite. In such cases, the base-satellite is regarded as the reference spacecraft. Because of the complexity of the problem, no thrusters on the sub-satellite are considered, and the effect of atmospheric drag, Earh's oblateness, and electrodynamic force on the spacecraft are neglected. Under these assumptions, Misra et al. \cite{Misra:2001} and Misra \cite{Misra:2008} showed that, in the station keeping phase (when the tether length is constant), the pitch motion (planar motion) of a TSS in the orbital plane is similar to that of a simple pendulum. The local vertical is a stable equilibrium position, which rotates uniformly about the orbit normal, and the local horizontal is an unstable equilibrium position. For the coupled pitch-roll motion, however, regions of quasi-periodic and chaotic motion are observed using numerical tools such as phase portraits and Poincar\'e sections \cite{Misra:2001, Misra:2008}. The equilibrium positions along the local vertical, local horizontal, and orbit normal do not disappear. Actually, the local vertical is commonly used in planar models as an initial condition for fast deployment/retrieval missions \cite{Vadali:1991, Pradeep:1997, Sun:2014}. This paper introduces the out-of-plane motion in the local vertical and local horizontal positions, such that the initial states fall in the chaotic zone during the state-keeping phase. Thus, a tension force in the tether is applied to extend the tether until a maximum length and to suppress the chaotic behavior of the TSS. A linear approximation about the vertical position has been used, which leads to a linear TCL for tether deployment \cite{Pradeep:1997, Kumar:1998, Sun:2014}. The simulations show that the tension law performs well and the chaotic motion can be guided to the stable vertical position. However, terminal oscillations of the roll angle are encountered during deployment.

\section{Equations of Motion of TSS} \label{sec:2}
\begin{figure}[hbt!]
\centering
\includegraphics[width=.8\textwidth]{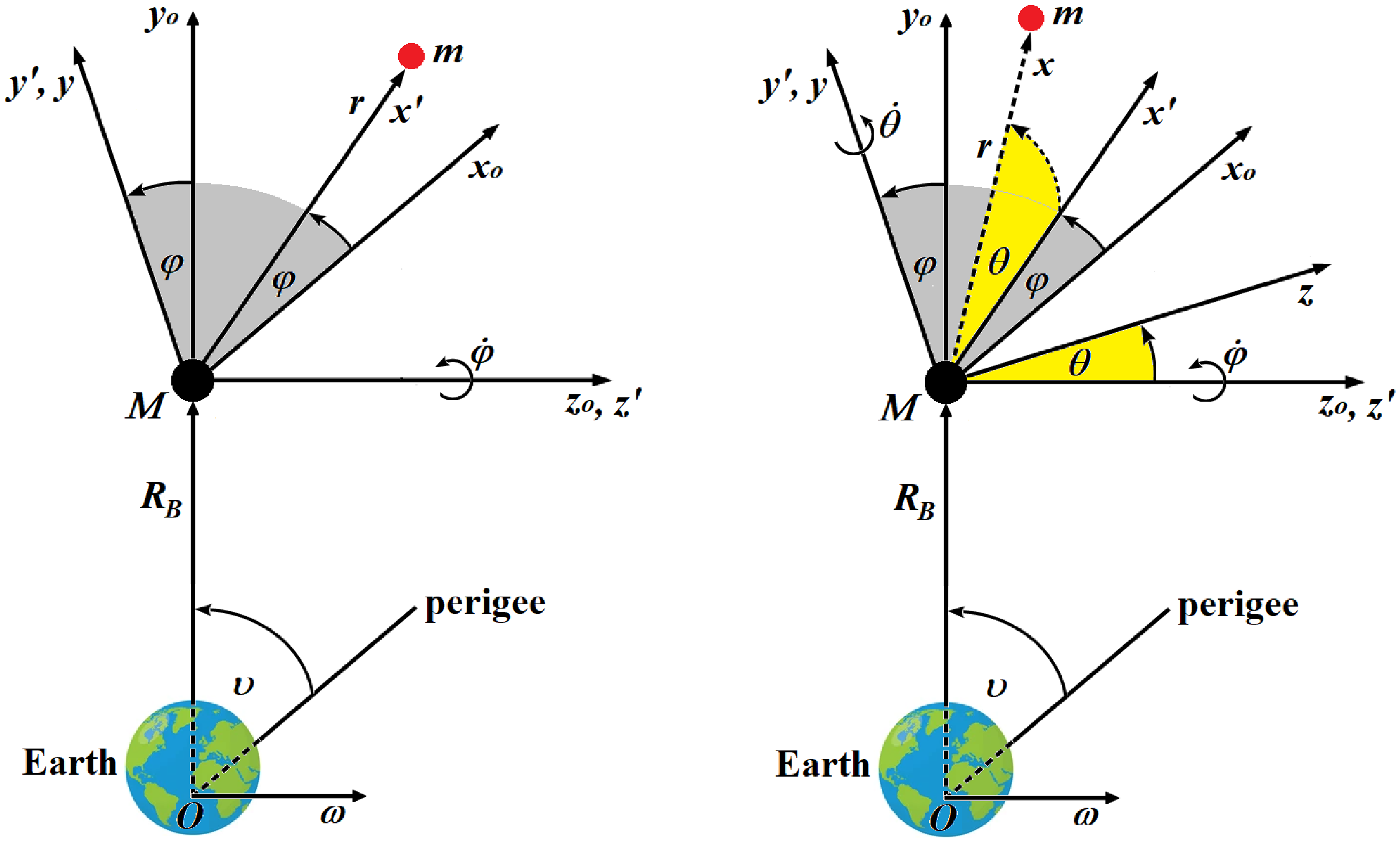}
\caption{\label{fig:1}Sketch of the geometry of TSS around the Earth showing reference coordinate systems and the spherical coordinate system. Pitch motion (left) is obtained first by rotating the $x_0y_0$ plane counterclockwise through an angle $\varphi$ with respect to the $z_0$ axis. Roll motion (right) is then obtained by rotating the $x'z'$ plane counterclockwise through an angle $\theta$ with respect to the $y'$ axis.}
\end{figure}

Figure \ref{fig:1} shows a schematic diagram of a TSS around the Earth. The system considered here consists of two spacecraft, a base-satellite; and a far smaller sub-satellite at the end of a massless and inextensible tether which is capable of being deployed from the base-satellite. The masses of the sub-satellite and the base-satellite are $m$ and $M$ ($m<<M$), respectively. The base-satellite is assumed to be in a Keplerian circular orbit with true anomaly $\upsilon$, orbital angular velocity $\omega$, and orbital radius $R_B$ from the Earth's center $O$. In such cases, the base-satellite is regarded as the reference spacecraft; and its orbital coordinate system centered at base-satellite, and designated by $x_0$-$y_0$-$z_0$ axes in Fig. \ref{fig:1}, as the reference coordinate system, such that, axis $y_0$ is directed along the position vector $R_B$ (i.e. aligned with the local upward vertical) and $z_0$ is aligned with the vector of the orbital angular velocity $\omega$, as shown in Fig. \ref{fig:1}. In this system, the position of the sub-satellite is represented by the radius vector $r$ in Fig. \ref{fig:1}, in such a way that its magnitude $r << R_B$. Therefore, the relative motion of the sub-satellite can be described by the variation of $r$. 

Using spherical coordinates, the radius vector $r$ can be described by its magnitude $r$, and the two direction angles $\varphi$ and $\theta$ (see Fig. \ref{fig:1}), where $\theta$ is the angle between $r$ and the plane $x_0$-$y_0$ (out-of-plane roll angle), and $\varphi$ is the angle between the projection of $r$ on the plane $x_0$-$y_0$ and the axis $x_0$ (in-plane pitch angle). The pitch-roll motion of the sub-satellite can be obtained by rotating $x_0$-$y_0$-$z_0$ axes through the angles $\varphi$ and $\theta$, respectively. The pitch motion is performed first by rotating the $x_0y_0$ plane counterclockwise through an angle $\varphi$ with respect to the $z_0$ axis, obtaining the coordinate system $x'$-$y'$-$z'$, as shown in Fig. \ref{fig:1} (left). The roll motion is then obtained by rotating the $x'z'$ plane counterclockwise through an angle $\theta$ with respect to the $y'$ axis, obtaining the coordinate system $x$-$y$-$z$, as shown in Fig. \ref{fig:1} (right). Additionally, $x$-axis is aligned with the vector $r$. On the other hand, we assume that the tether remains straight during deployment, then the instantaneous tether length $\ell$ is equal to the distance between the base-satellite and the sub-satellite, i.e. $r = \ell$. In this manner, the equations of motion that describe the relative motion between the two spacecraft according to the assumptions made in this study, are given by \cite{Yu:2000}:
\begin{subequations}
\begin{align}
\label{eq:1a}
\ddot{\ell} - \ell\dot{\theta}^2 - \ell\left(\omega + \dot{\varphi}\right)^2\cos^2{\theta} - \ell\omega^2\left(3\sin^2{\varphi}\cos^2{\theta}-1\right) &=  -\frac{\mathcal{T}}{m},\\
\label{eq:1b}
\ddot{\varphi} + 2\frac{\dot{\ell}}{\ell}\left(\omega + \dot{\varphi}\right) - 2\left(\omega + \dot{\varphi}\right)\dot{\theta}\tan{\theta} - 3\omega^2\sin{\varphi}\cos{\varphi} &=  0,\\
\label{eq:1c}
\ddot{\theta} + 2\frac{\dot{\ell}}{\ell}\dot{\theta} + \left(\omega + \dot{\varphi}\right)^2\cos{\theta}\sin{\theta} + 3\omega^2\sin^2{\varphi}\sin{\theta}\cos{\theta} &=  0,
\end{align}
\end{subequations}
where $\mathcal{T}$ is the tension tether force, and overdot denotes differentiation with respect to time $t$. Equation \eqref{eq:1a} describes the variation of the tether length, and Eqs. \eqref{eq:1b} and \eqref{eq:1c} describe the pitch-roll motion of the sub-satellite, respectively. These equations can be nondimensionalized by using the following nondimensional variables:
\begin{eqnarray} 
\upsilon &=& \omega t,\;\;\;\; \rho = \frac{\ell}{L},\;\;\;\; u = \frac{\mathcal{T}}{m\omega^2L},\nonumber   
\end{eqnarray}     
where $L$ is the maximum tether length. The nondimensional equations are given by the following:
\begin{subequations}
\begin{align}
\label{eq:2a}
\rho'' - \rho\theta'^2 - \rho\left(1 + \varphi'\right)^2\cos^2{\theta} - \rho\left(3\sin^2{\varphi}\cos^2{\theta}-1\right) &=  -u,\\
\label{eq:2b}
\varphi'' + 2\frac{\rho'}{\rho}\left(1 + \varphi'\right) - 2\left(1 + \varphi'\right)\theta'\tan{\theta} - 3\sin{\varphi}\cos{\varphi} &=  0,\\
\label{eq:2c}
\theta'' + 2\frac{\rho'}{\rho}\theta' + \left(1 + \varphi'\right)^2\cos{\theta}\sin{\theta} + 3\sin^2{\varphi}\sin{\theta}\cos{\theta} &=  0,
\end{align}
\end{subequations}
where prime refers to differentiation with respect to the true anomaly $\upsilon$. 

In the station keeping phase (when the tether length is constant) $\ell' = 0$. Then Eqs. \eqref{eq:2a} to \eqref{eq:2c} reduce to
\begin{subequations}
\begin{align}
\label{eq:3a}
\varphi'' - 2\left(1 + \varphi'\right)\theta'\tan{\theta} - 3\sin{\varphi}\cos{\varphi} &=  0,\\
\label{eq:3b}
\theta'' + 2\left(1 + \varphi'\right)^2\cos{\theta}\sin{\theta} + 3\sin^2{\varphi}\sin{\theta}\cos{\theta} &=  0.
\end{align}
\end{subequations}
This system has three equilibrium points $\left(\varphi_e,\varphi'_e,\theta_e, \theta'_e\right)^T$: $\left(\pm(2n+1)\frac{\pi}{2},0,\pm n\pi,0\right)^T$ (the local vertical), $\left(\pm n\pi,0,\pm n\pi,0\right)^T$ (the local horizontal), and $\left(\varphi^{*}, 0,\pm(2n+1)\frac{\pi}{2},0\right)^T$ (the orbit normal), where $n$ is any integer and $\varphi^{*}$ is any constant. Only the local vertical position is stable \cite{Misra:2001, Misra:2008}. Note that, if the roll motion is initially unexcited ($\theta_0 = \theta'_0 = 0$), the motion of the TSS is confined to the orbital plane, and described by
\begin{equation}
\label{eq:4}
\varphi'' - 3\sin{\varphi}\cos{\varphi} =  0.
\end{equation}

Multiplying Eq. \eqref{eq:4} by $\varphi'$ and integrating it, it is easy to find a nondimensional integral $E$ (i.e. $E' = 0$) that takes on the form 
\begin{equation}
\label{eq:5}
E = \varphi'^2 + 3\cos^2{\varphi}.
\end{equation}

\begin{figure}[hbt!]
\centering
\includegraphics[width=.5\textwidth]{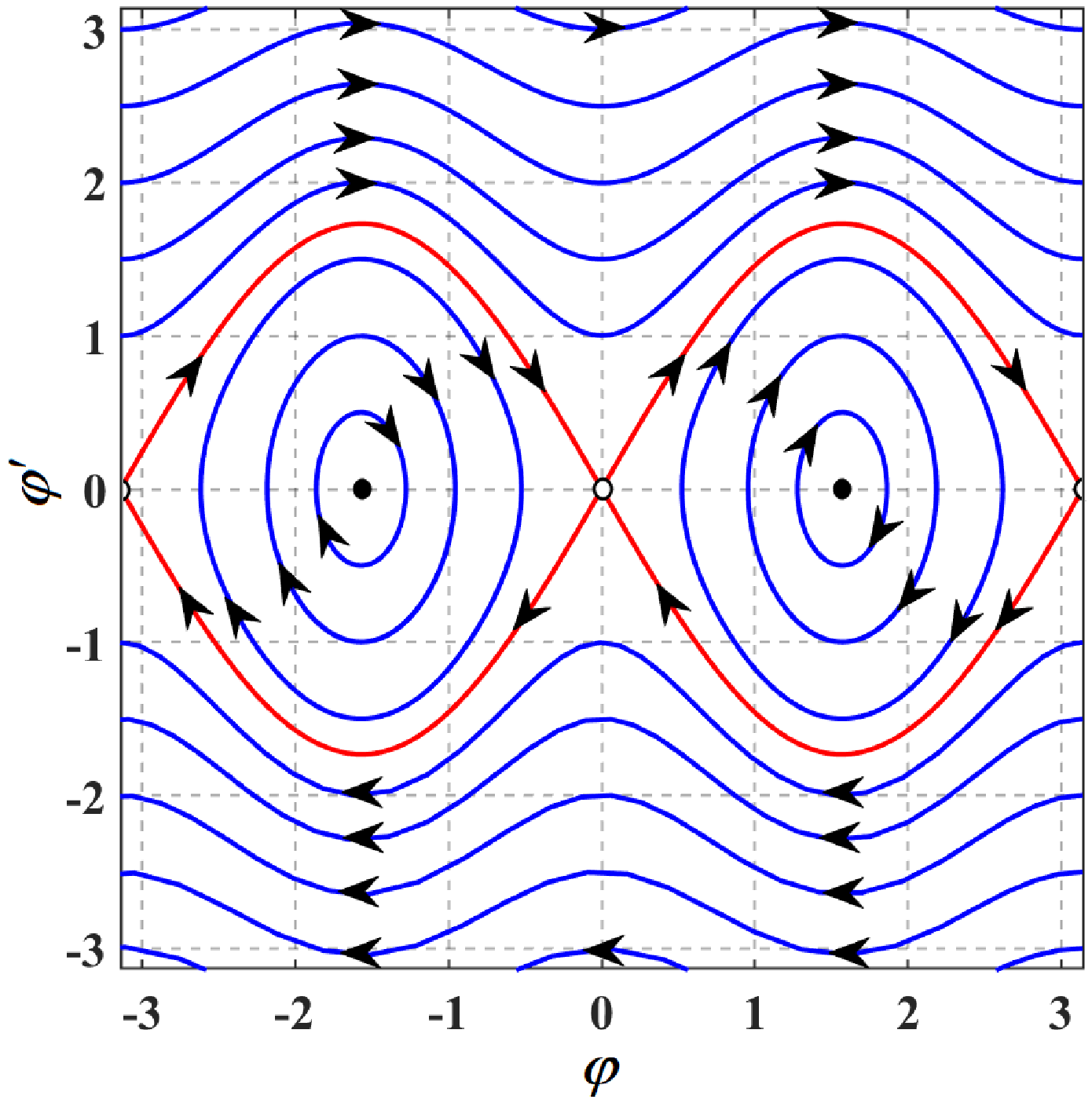}
\caption{\label{fig:2}Phase portrait of in-plane pitch motion.}
\end{figure}

Figure \ref{fig:2} shows the phase portrait of Eq. \eqref{eq:4}. If the length of the tether does not vary, the dynamics of the in-plane motion of the TSS is similar to one valid for a simple pendulum. The stable points $\left(\pm(2n+1)\frac{\pi}{2},0,\pm n\pi,0\right)^T$ ($E = 0$) correspond to centers, with the local upward and downward vertical positions at rest and small oscillations (closed orbits with $0 < E < 3$ ) about centers. On other hand, the unstable points $\left(\pm n\pi,0,\pm n\pi,0\right)^T$ ($E = 3$) correspond to saddles, with heteroclinic trajectories joining the equilibrium points. Therefore, the local horizontal positions represent an inverted pendulum at rest. For $E > 3$, the TSS whirls repeatedly over the orbital plane, as shown in Fig. \ref{fig:2}.          

\section{Analysis of Chaotic Motion} \label{sec:3}
In the last section, the phase portrait of planar motion showed that the in-plane dynamics of the TSS is similar to the one used for a simple pendulum. However, if roll motion is initially excited (i.e. $\theta_0 \neq 0$ or $\theta'_0 \neq 0$), then the three-dimensional coupled pitch-roll motion of the TSS becomes more complex. Firstly, the phase space has four dimensions: $\varphi$, $\varphi'$, $\theta$, $\theta'$. Similarly, there exist a nondimensional integral $C$ (i.e. $C' = 0$) for the coupled system that can be written as \cite{Misra:2001, Misra:2008}
\begin{equation}
\label{eq:6}
C = \theta'^2 + \cos^2{\theta}\left(\varphi'^2 - 1 - 3\sin^2{\varphi}\right).
\end{equation}

\begin{figure}[hbt!]
\centering
\includegraphics[width=1.0\textwidth]{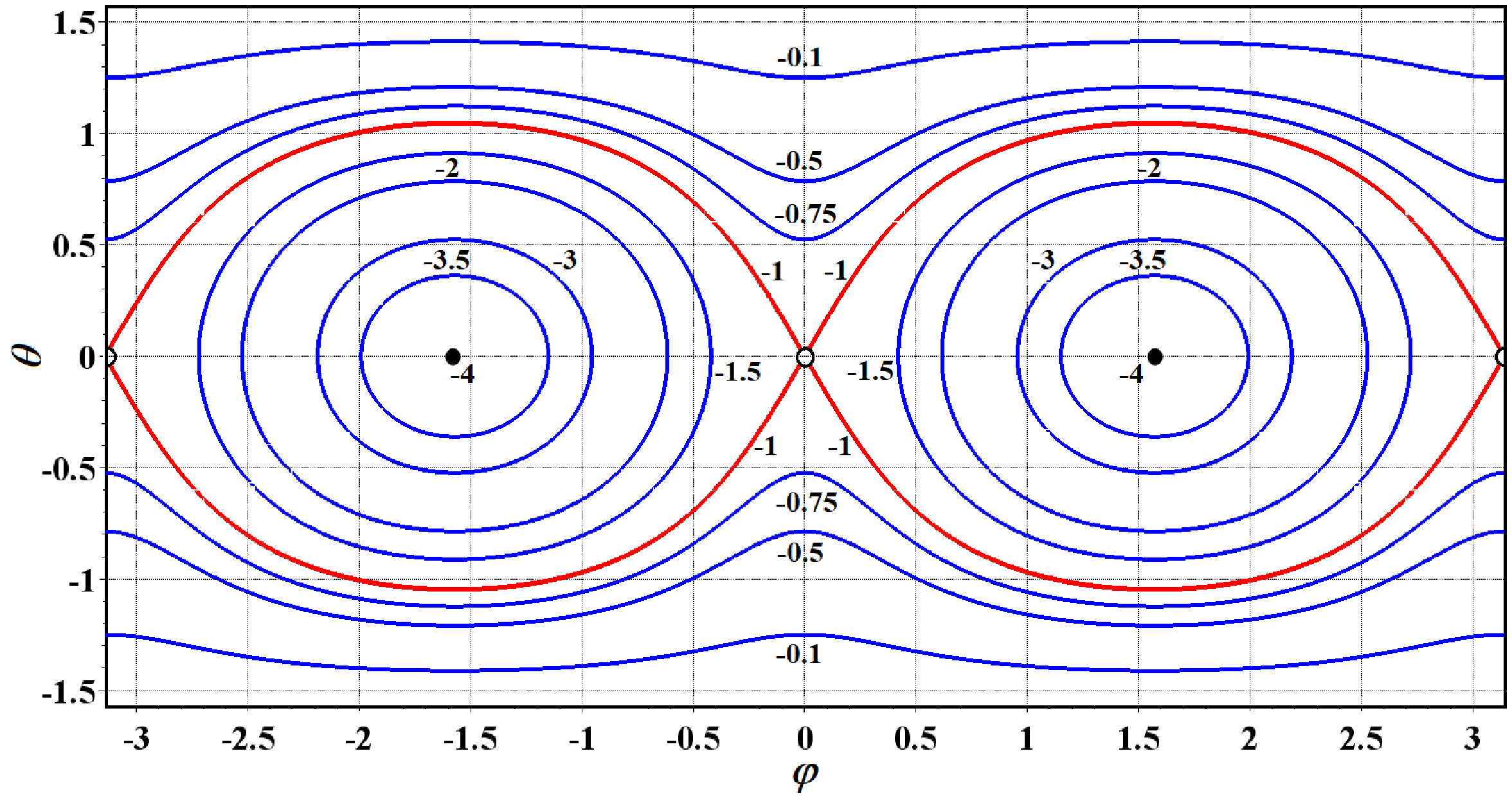}
\caption{\label{fig:3}Zero velocities curves in $\varphi$-$\theta$ space.}
\end{figure}

Setting $\varphi' = \theta' = 0$ for a given constant $C$, Eq. \eqref{eq:6} provides an algebraic expression for the zero velocity curves in $\varphi-\theta$ space. Selected zero velocity curves are shown in Fig. \ref{fig:3}. These curves determine the geometric extremes of possible motion for a given constant $C$. For example, the minimum value for $C$ occurs in the local vertical position when $C = -4$. Therefore, no motion is possible when $C < -4$; for $-4\leq C \leq -1$ motion is bounded in both $\varphi$ and $\theta$. Note that $C = -1$ in the local horizontal position. Finally, for $-1\leq C \leq 0$ the motion is bounded in $\theta$ only.                

\begin{figure}[hbt!]
\centering
\includegraphics[width=1.0\textwidth]{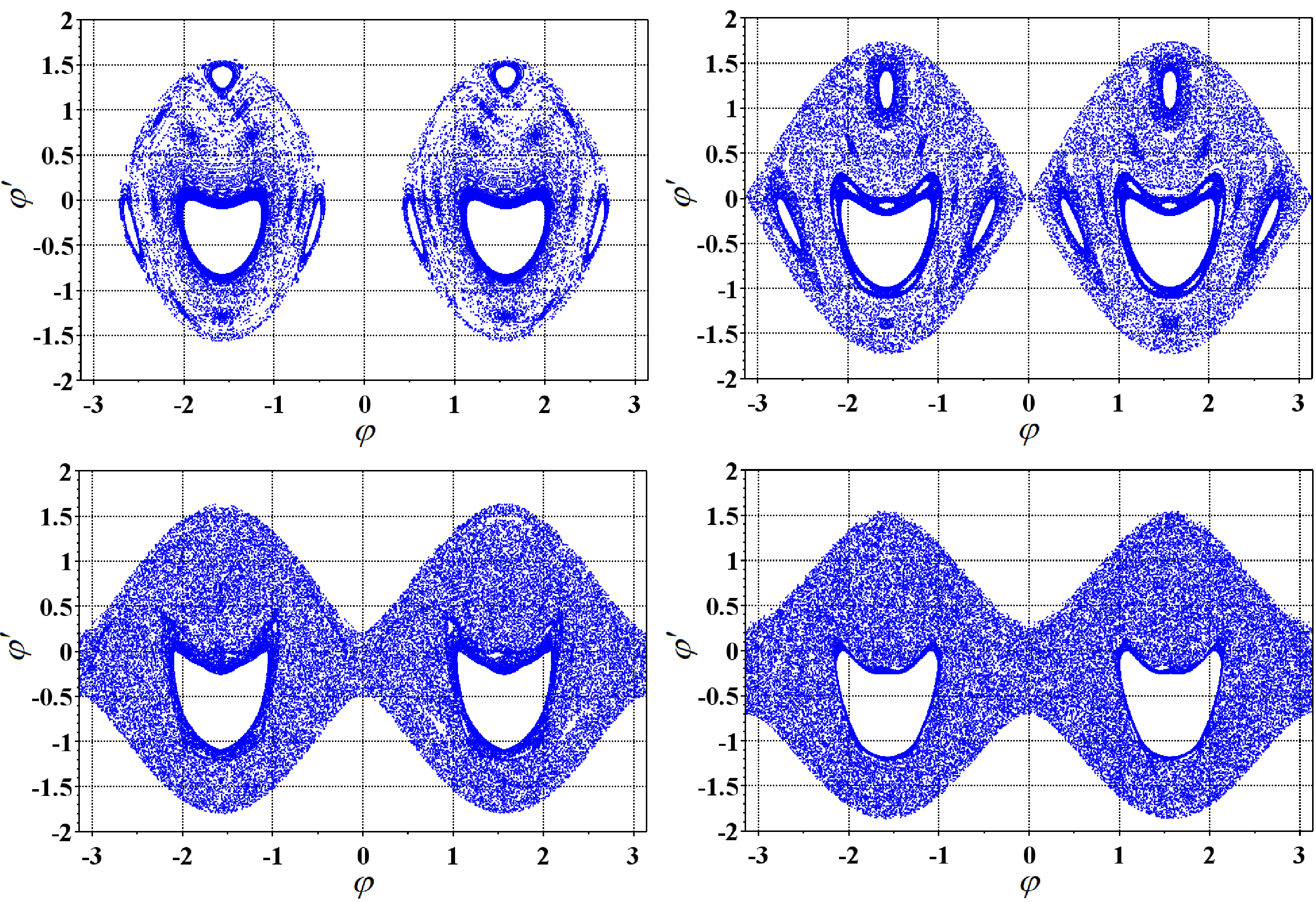}
\caption{\label{fig:4}Poincar\'e sections for coupled motion for various $C > -4$. $C = -1.5$ (top left), $C = -1.0$ (top right), $C = -0.75$ (bottom left), and $C = -0.5$ (bottom right) are shown.}
\end{figure}

Because the phase space has four dimensions and the system formed by Eqs. \eqref{eq:3a} and \eqref{eq:3b} has an integral, the coupled motion of the TSS may be studied by making a Poincar\'e section \cite{Moon:1992}. For a given value of $C$, Eqs. \eqref{eq:3a} and \eqref{eq:3b} are integrated numerically with different initial conditions to produce several trajectories with the same value of $C$. The two dimensional Poincar\'e map of the phase trajectories is taken as $\Sigma = \left\{\left(\varphi,\varphi'\right)| \theta = 0, \theta' > 0\right\}$, sampled at period $2\pi$. Figure \ref{fig:4} shows the Poincar\'e sections for $C = -1.5$, $C = -1.0$, $C = -0.75$, $C = -0.25$, respectively. Each figure is made up of several trajectories with $\varphi_0 = \varphi^*$, $\varphi'_0 = 0$, $\theta = 0$, $\theta' = \sqrt{C+ \left(1 + 3\sin^2{\varphi_0}\right)}$, for various values of $\varphi^*$, over 500 orbits. In these Poincar\'e sections, there are islands of regular motion (i.e. periodic and quasi-periodic solutions); however, chaotic trajectories appear as a cloud of disorganized points for $C > -1$ (Fig. \ref{fig:4} bottom) and the size of the chaotic region increases as the value of $C$ is increased. 

\begin{figure}[hbt!]
\centering
\includegraphics[width=0.475\textwidth]{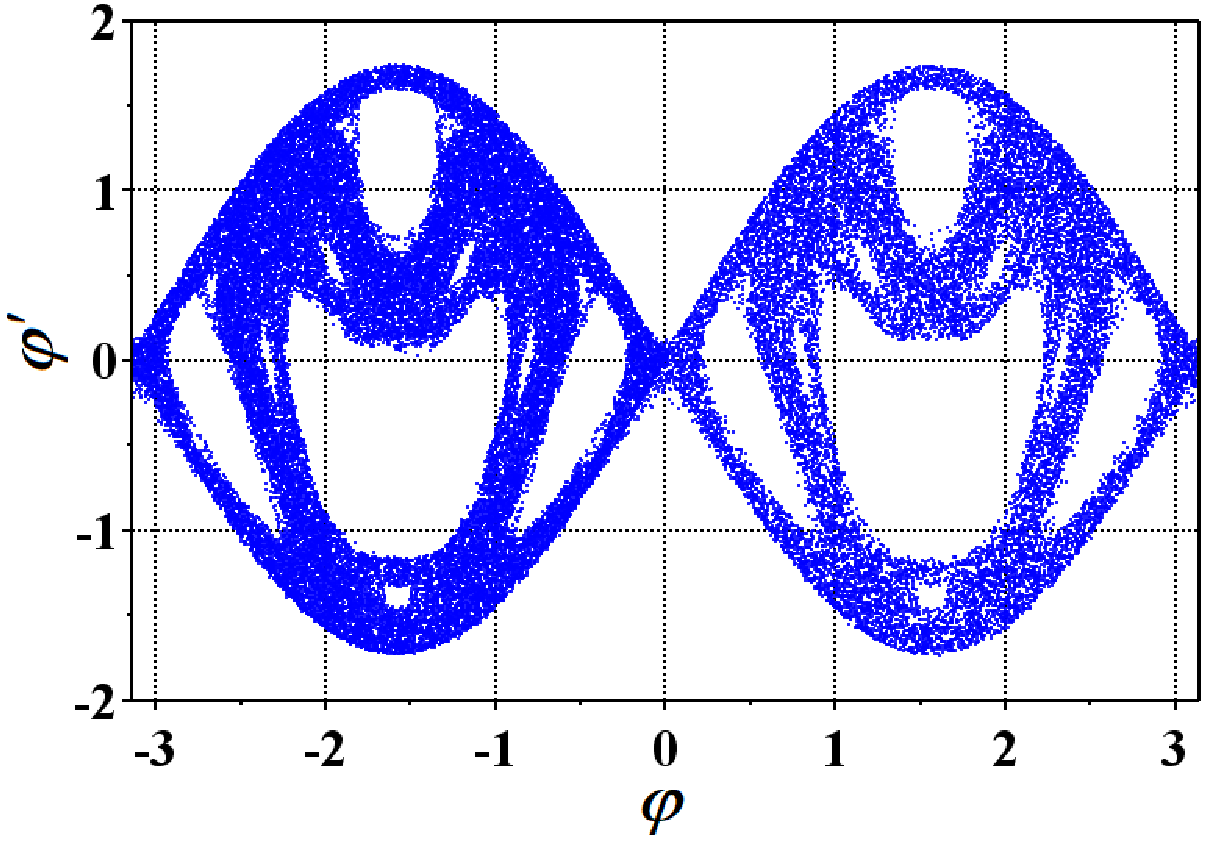}
\caption{\label{fig:5}Poincar\'e sections for coupled motion with $\varphi_0 = 0$, $\theta_0 = \theta^*$, $\varphi'_0 = \theta'_0 = 0$, where $0\leq\theta^* \leq \pi/12$.}
\end{figure}

In order to study the control of a three dimensional chaotic motion of the TSS, the system is considered to begin with initial states that fall in the chaotic zone. Thus, two equilibrium positions have been chosen: the local vertical $\left(\frac{\pi}{2},0,0,0\right)^T$ and the local horizontal $\left(0,0,0,0\right)^T$. The first one is commonly used as an initial condition for fast deployment/retrieval missions. The local vertical position will be excited introducing the roll motion $\theta_0' = \sqrt{C+ \left(1 + 3\sin^2{\varphi_0}\right)}$ for $C > -1$, producing an initial state where chaos happens (see Fig. \ref{fig:4}). On the other hand, if $\varphi_0 = 0$ and $\theta_0 \neq 0$, then $C > -1$ (see Fig. \ref{fig:3}), and chaos might occur in the neighborhood of the local horizontal state $\left(0,0,0,0\right)^T$, as shown in Fig. \ref{fig:5}, where the system is now examined by applying the initial state $\varphi_0 = 0$, $\varphi_0' = 0$, $\theta_0 = \theta^*$, $\theta_0' = 0$ near one of the saddle point $\left(0,0,0,0\right)^T$, for various values of $\theta^*$. Thus, chaos can occur from the local horizontal position by introducing the roll angle $\theta_0$. This work is then concerned with perturbed local vertical and local horizontal positions as initial states such that chaotic motion exists.

\section{Tether length control} \label{sec:4}
This section describes the linear scheme used to determine the tension force in the tether that would suppress the chaotic motion mentioned above. This approach proposes a linearization about a stable equilibum point. Our goal is to initiate the TSS in the chaotic zone with $\rho_0 = 0$ and to extend the tether until a maximum length by varying the tension $u$ in the tether, such that the system will achieve the desired stable state $\left(\rho_f, \rho_f', \varphi_f, \varphi_f', \theta_f, \theta_f'\right)^T = \left(1,0,\pm \frac{\pi}{2},0,0,0\right)^T$.         

Assuming that perturbations about the final state $\left(\rho_f, \rho_f', \varphi_f, \varphi_f', \theta_f, \theta_f'\right)^T = \left(1,0,\pm \frac{\pi}{2},0,0,0\right)^T$ are small, and that the nondimensional TCL $u$ is a linear function of the state variables, then Eqs. \eqref{eq:2a} to \eqref{eq:2c} can be linearized about $\left(1,0,\pm \frac{\pi}{2},0,0,0\right)^T$. It is possible to define the new variables $x_i$, so that $\rho = 1 + x_1$, $\rho' = x_2$, $\varphi = \pm\frac{\pi}{2} + x_3$, $\varphi' = x_4$, $\theta = x_5$, $\theta' = x_6$. If nonlinearities were sufficiently small in magnitude, then the linearized equations of the system can be obtained from Eqs. \eqref{eq:2a} to \eqref{eq:2c}, as follows:

\begin{subequations}
\begin{align}
\label{eq:7a}
x_1''  - 3x_1 - 2x_3' &= 3 - u,\\
\label{eq:7b}
x_3''  + 2x_1' + 3x_3 &= 0,\\
\label{eq:7c}
x_5''   + 4x_5 &= 0,
\end{align}
\end{subequations}
where $u$ is assumed to be a linear function of the state variables, that is
\begin{equation}
\label{eq:8}
u = \sum_{i=1}^{6}K_ix_i + K_7.
\end{equation}
Combining Eqs. \eqref{eq:7a} and \eqref{eq:8}, the linear system is transformed to  
          
\begin{equation}
\label{eq:9}
\begin{pmatrix}
x''_1 \\
x''_3 \\
x''_5
\end{pmatrix} + 
\begin{pmatrix}
K_2 & -2+K_4 & K_6 \\
2   &   0    & 0 \\
0   &   0    & 0
\end{pmatrix} 
\begin{pmatrix}
x'_1 \\
x'_3 \\
x'_5
\end{pmatrix} + 
\begin{pmatrix}
-3 + K_1 & K_3    & K_5 \\
0        &   3    & 0 \\
0        &   0    & 4
\end{pmatrix} 
\begin{pmatrix}
x_1 \\
x_3 \\
x_5
\end{pmatrix} 
=
\begin{pmatrix}
3 - K_7 \\
0 \\
0
\end{pmatrix}.
\end{equation}
Let $K_7 = 3$. This makes the linear second order equation \eqref{eq:9} homogeneous, and therefore the final state becomes in a zero solution for Eq. \eqref{eq:9}. To investigate the stability of the zero solution, Eq. \eqref{eq:9} is now of the form
\begin{equation}
\label{eq:10}
q'' + Gq' + Kq  = 0,
\end{equation}
where
\begin{eqnarray} 
q &=& 
\begin{pmatrix}
x_1 \\
x_2 \\
x_3
\end{pmatrix},\;\;\;\; 
G = \begin{pmatrix}
K_2 & -2+K_4 & K_6 \\
2   &   0    & 0 \\
0   &   0    & 0
\end{pmatrix},\;\;\;\; 
K = \begin{pmatrix}
-3 + K_1 & K_3    & K_5 \\
0        &   3    & 0 \\
0        &   0    & 4
\end{pmatrix}.\nonumber   
\end{eqnarray} 
According to the Zajac's extension of the KTC (Kelvin-Talt-Chetayev) theorem \cite{Pradeep:1997, Zajac:1964}, the zero solution of the equation $q''+ Gq' + Kq = 0$ is asymptotically stable if the equation $q''+ Kq = 0$ is stable and the matrix $D$ is positive semi-definite. Examining the entries of the matrices $G$ and $K$, equation $q''+ Kq = 0$ is stable if $K_1 > 3$, $K_3 = K_5 = 0$, and the matrix $D$ is positive semi-definite if $K_2 > 0$, $K_4 = K_6 = 0$. 

Based on the KTC theorem, the controlled linear system is asymptotically stable about $\left(\rho_f, \rho_f', \varphi_f, \varphi_f', \theta_f, \theta_f'\right)^T = \left(1,0,\pm \frac{\pi}{2},0,0,0\right)^T$ if the tension is             
\begin{equation}
\label{eq:11}
u = K_1\left(\rho - 1\right) + K_2\rho' + 3,
\end{equation}
with $K_1 > 3$, $K_2 >0$. This control law is simple because it only requires to measure the tether deployed length and its velocity, and to adjust the tension, which settles down to the equilibrium value of $3$.     

In what follows, we consider the deployment of a TSS traveling in a circular orbit at an altitude of $220$ km, with $\omega = 1.1804 \times 10^{-3}$ rad/s. The tether is assumed to be $1$ km long ($L = 1$ km). The masses of the satellites are set as $M = 1000$ kg and $m = 50$ kg.     

\section{Numerical Results} \label{sec:5}
\begin{figure}[hbt!]
\centering
\includegraphics[width=1.0\textwidth]{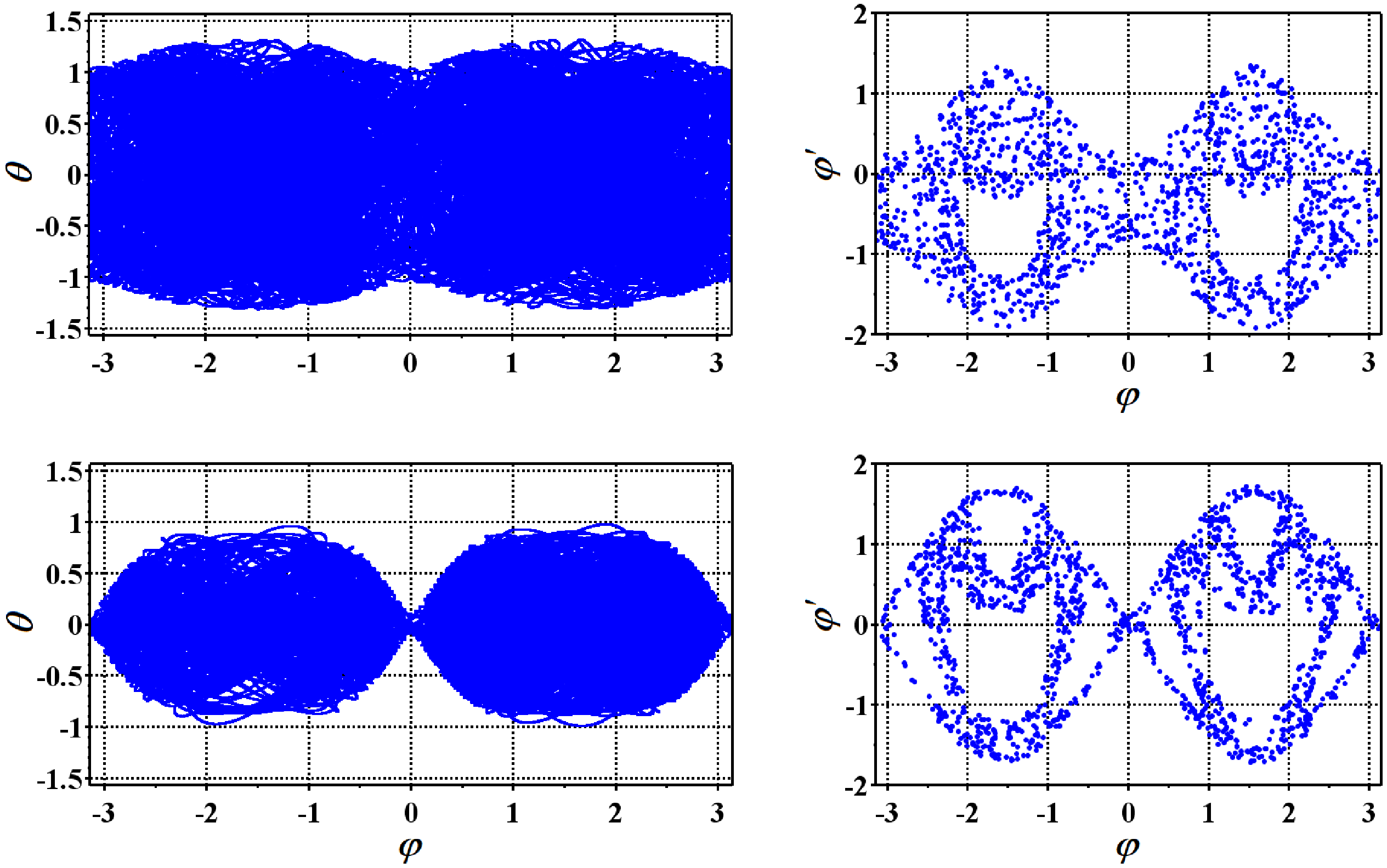}
\caption{\label{fig:6}Chaotic motion (left) and Poincar\'e section (right) of three trajectories with $\varphi_0 = \frac{\pi}{2}$, $\varphi_0' = 0$, $\theta_0 = 0$, $\theta_0' = \sqrt{3.75}$ (up) and $\varphi_0 = 0$, $\varphi_0' = 0$, $\theta_0 = \frac{\pi}{30}$, $\theta_0' = 0$ (bottom). Constant length case.}
\end{figure}

To show and control the chaotic motion of the TSS, the system is considered to begin with the following initial states $\left(\rho_0, \rho'_0,\varphi_0,\varphi'_0,\theta_0,\theta'_0\right)^T = \left(0.01,0.5,\frac{\pi}{2},0,0,\sqrt{C+4}\right)^T$ and $\left(\rho_0, \rho'_0,\varphi_0,\varphi'_0,\theta_0,\theta'_0\right)^T = \left(0.01,0.5,0,0,\frac{\pi}{30},0\right)^T$. The initial condition $\rho_0 = 0.01$ ($\ell_0 = 0.01$ km) is selected because $\rho_0 = 0$ creates a singularity in Eqs. \eqref{eq:2b} and \eqref{eq:2c}, and $\rho'_0 = 0.5$ ($\dot{\ell}_0 = 0.59$ m/s) is the same value used in \cite{Vadali:1991, Pradeep:1997, Sun:2014}. The initial pitch and roll motion conditions were set such that they fall in the chaotic zone during the station-keeping phase, as shown in Figs. \ref{fig:6} (right), where Fig. \ref{fig:6} (up) corresponds to $\varphi_0 = \frac{\pi}{2}$, $\varphi'_0 = 0$, $\theta_0 = 0$, $\theta'_0 = \sqrt{C+4}$, with $C = -0.25$ ($\dot{\theta}_0 = 2.29\times 10^{-3}$ rad/s), and Fig. \ref{fig:6} (bottom) corresponds to $\varphi_0 = 0$, $\varphi'_0 = 0$, $\theta_0 = \frac{\pi}{30}$, $\theta'_0 = 0$. The Poincar\'e sections obtained from these trajectories, and shown in Fig. \ref{fig:6} (right), ensure the existence of chaos. The effect of the tether length control laws derived above and proposed on chaos is evaluated by solving numerically the nonlinear equations \eqref{eq:2a} to \eqref{eq:2c}. 

To guide the chaotic motion observed in Fig. \ref{fig:6} to the desired final state $\left(\rho_f, \rho_f', \varphi_f, \varphi_f', \theta_f, \theta_f'\right)^T = \left(1,0,\pm \frac{\pi}{2},0,0,0\right)^T$, Eqs. \eqref{eq:2a} to \eqref{eq:2c} are solved numerically using the linear TCL $u$ given by Eq. \eqref{eq:11}. Based on the stability condition for $K_1$ and $K_2$ of the systems described above, the gains selected are $K_1 = 4$, $K_2 = 3$ for the linear tension \eqref{eq:11}. Finally, it is important to mention that the tension $u$ must be positive, i.e. $u \geq 0$, at any time to confirm the workability of the control law. In the numerical simulations performed in this work, the tension is always positive. 

\begin{figure}[hbt!]
\centering
\includegraphics[width=0.95\textwidth]{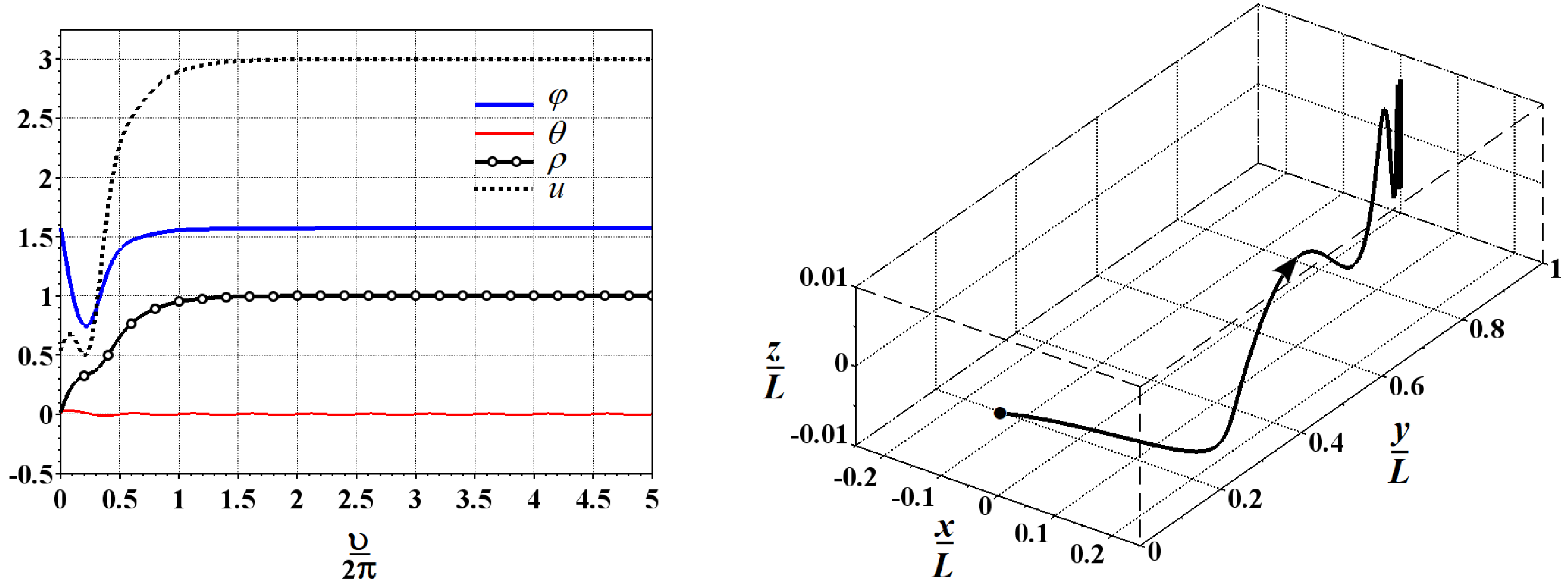}
\caption{\label{fig:7}The effect of the deployment on chaotic trajectories (left) and three-dimensional subsatellite's trajectory (right) using the linear control law. The system initiates with $\varphi_0 = \frac{\pi}{2}$, $\varphi'_0 = 0$, $\theta_0 = 0$, $\theta_0' = \sqrt{3.75}$.}
\end{figure}

\begin{figure}[hbt!]
\centering
\includegraphics[width=0.95\textwidth]{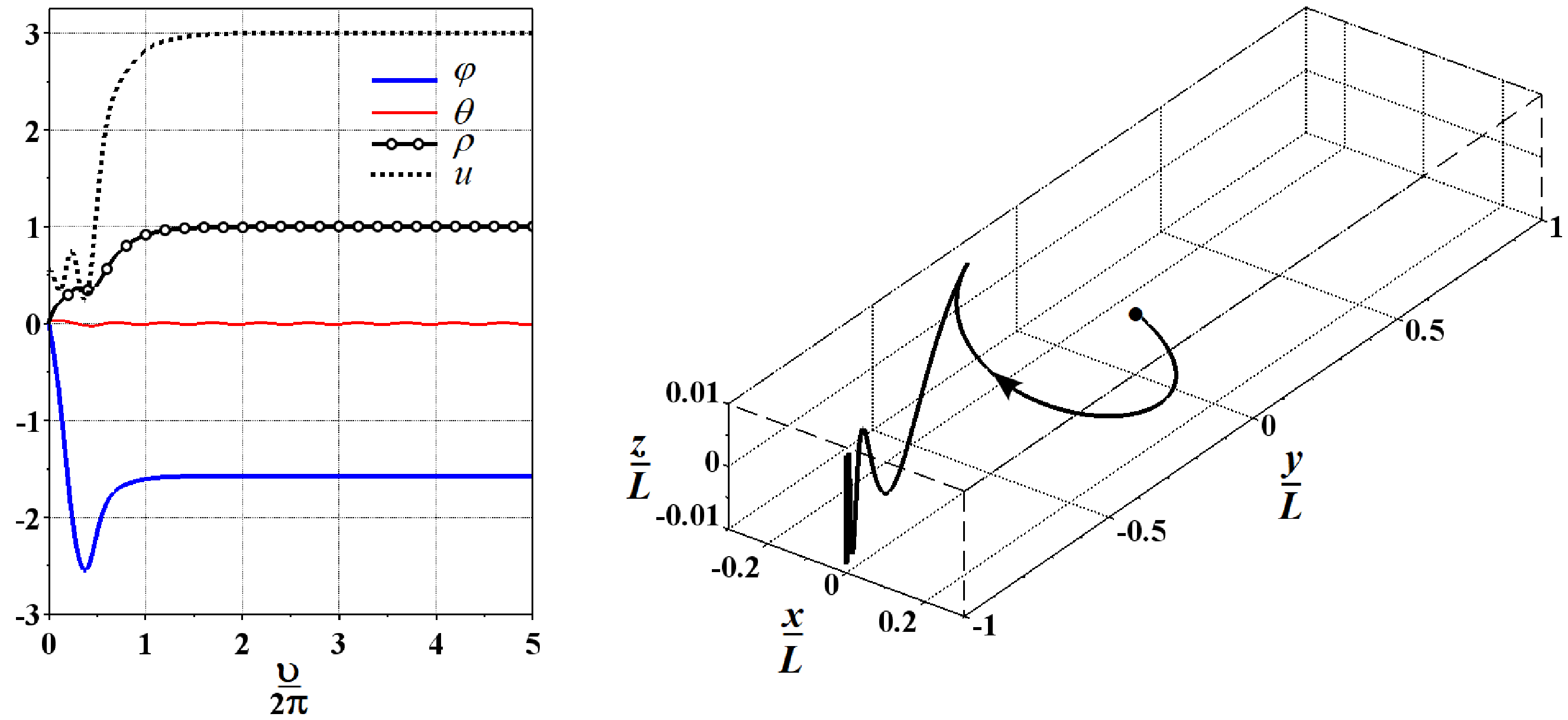}
\caption{\label{fig:8}The effect of the deployment on chaotic trajectories (left) and three-dimensional subsatellite's trajectory (right) using the linear control law. The system initiates with $\varphi_0 = 0$, $\varphi'_0 = 0$, $\theta_0 = \frac{\pi}{30}$, $\theta_0' = 0$.}
\end{figure}

Figures \ref{fig:7} and \ref{fig:8} show the effect of the TCL on the chaotic trajectories depicted in Figs. \ref{fig:6} (top) and \ref{fig:6} (bottom), respectively. The plots of the nondimensional tether length, the pitch-roll angles, and the nondimensional tension control versus true anomaly, $\upsilon$, are shown in Figs. \ref{fig:7} (left) and \ref{fig:8} (left). It can be seen that the linear control \eqref{eq:11} completes the deployment phase in nearly two orbits. No tether length oscillations are observed and the length settles down to $\rho = 1$. Additionally, the controlled results in Figs. \ref{fig:7} (left) and \ref{fig:8} (left) show that the nondimensional tension is smooth, and it performs well and achieves the equilibrium value of $3$ ($\tau = 0.21$ N); however, small oscillations of the roll angle are encountered at the ending of deployment, as shown in Figs. \ref{fig:7} (left) and \ref{fig:8} (left). Note that the tether dynamics shown in Eqs. \eqref{eq:2a} to \eqref{eq:2c} is an underactuated system. These oscillations actually behaves like a quasi-periodic motion about the equilibrium point, as shown in Figs. \ref{fig:7} (right) and \ref{fig:8} (right).   

Figures \ref{fig:7} (right) and \ref{fig:8} (right) show the three-dimensional sub-satellite's trajectory being guided to the stationary point $\left(\frac{x}{L},\frac{y}{L},\frac{z}{L}\right)^T = \left(0,\pm 1,0\right)^T$ that corresponds to the terminal phase of deployment. Note that the scales on the axes $x_0$, $y_0$, $z_0$ are not the same to visualize the three-dimensional motion. The coordinates of the sub-satellite in the $x_0$-$y_0$-$z_0$ orbital system can be determined by  $x = \ell\cos{\theta}\cos{\varphi}$, $y = \ell\cos{\theta}\sin{\varphi}$, $z = -\ell\sin{\theta}$. As observed from Figs. \ref{fig:7} (right) and \ref{fig:8} (right), no chaotic motion exists and the resulting TSS motion path is smooth and fast, which is achieved through tether length control laws outlined in this work. As was mentioned above, the subsatellite is guided from the chaotic motion in Fig. \ref{fig:6} to a quasi-periodic motion that has small oscillations about the local vertical, in which the tether satellite holds a length constant of $1$ km.                                              

\section{Conclusions} \label{sec:6}
In the station-keeping phase of the simplified model of a TSS in a circular orbit, if the motion of the TSS is assumed to be in the orbital plane, the dynamics is similar to that of a simple pendulum, such that the local vertical is a stable equilibrium position, and the local horizontal is an unstable equilibrium position. In the case of coupled motion, the dynamics of the TSS is more complex, and the Poincar\'e sections plots of the nonlinear dynamics show that the possible motion depends on the magnitude of the initial conditions. As initial conditions are increased, the motion changes from regular to chaotic. Actually, the local vertical, which is a common initial state for fast deployment of payloads, changes from stable equilibrium position to chaotic trajectory if roll motion is initially excited. Starting the TSS from perturbed vertical and horizontal initial positions, the dynamics and control of chaos on tethered satellites is studied in this paper.                

As a main contribution of the present research, a linear scheme has been presented for controlling chaos on tethered satellites. This approach linearized the equations of motion of the TSS about the equilibrium vertical position, and it leads to a linear tension. From the stability analysis, the desired final vertical position is asymptotically stable, but not globally stable. It is shown that a fast deployment can be accomplished by this scheme. It is observed that the TLC proposed in this work is capable of controlling the chaotic motion to the desired final state. The study shows that the perturbed vertical and horizontal positions lies within the domain of attraction. However, since the system is not globally stable, the orientation of the final state on the local vertical depends on the initial conditions. Finally, terminal small oscillations of the roll motion are encountered at the ending of the deployment phase. The result is small quasi-periodic orbits about the local vertical. A new control law will be designed to hold the row angle constant during the terminal phase of deployment by using more sophisticated models of a TSS.  



\section*{Acknowledgments}
The authors thank the reviewers for their constructive comments, which helped to improve the manuscript. This work has been supported by CAPES (Coordination for the Improvement of Higher Education Personnel, Brazil), Grant 88887.478205 /2020-00, and Fapesp (São Paulo Research Foundation, Brazil), Grant 2016/ 24561-0.   

\bibliography{sample.bib}

\end{document}